\documentclass[amsmath,amssymb,floatfix,preprintnumbers,nofootinbib,prd,twocolumn,superscriptaddress]{revtex4}

\usepackage[utf8]{inputenc}
\usepackage{graphicx}

\usepackage{amssymb}
\usepackage{amsxtra}
\usepackage{amsmath}
\usepackage{booktabs,multirow,tabularx}
\usepackage{slashed}
\usepackage{float}
\usepackage{placeins}
\usepackage{rotating}
\usepackage{lscape}
\usepackage{color}
\usepackage{hyperref}
\usepackage{cleveref}
\usepackage{cancel}
\usepackage[Q=yes,pverb-linebreak=no]{examplep}
\usepackage{ulem}

\DeclareGraphicsExtensions{.pdf}

\newcommand{\vev}[1]{\langle {#1} \rangle}

\newcommand{\gsim}{\gtrsim}

\def\beq{\begin{equation}}
\def\bea{\begin{eqnarray}}
\def\eeq{\end{equation}}
\def\eea{\end{eqnarray}}
\def\beqal{\begin{align}}
\def\endal{\end{align}}

\definecolor{applegreen}{rgb}{0.55, 0.71, 0.0}
\definecolor{purple}{rgb}{0.5,0.0,0.5}

\makeatletter
\newcommand\footnoteref[1]{\protected@xdef\@thefnmark{\ref{#1}}\@footnotemark}
\makeatother

\DeclareFontFamily{U}{cbgreek}{}
\DeclareFontShape{U}{cbgreek}{m}{n}{
        <-6>    grmn0500
        <6-7>   grmn0600
        <7-8>   grmn0700
        <8-9>   grmn0800
        <9-10>  grmn0900
        <10-12> grmn1000
        <12-17> grmn1200
        <17->   grmn1728
      }{}
\DeclareFontShape{U}{cbgreek}{bx}{n}{
        <-6>    grxn0500
        <6-7>   grxn0600
        <7-8>   grxn0700
        <8-9>   grxn0800
        <9-10>  grxn0900
        <10-12> grxn1000
        <12-17> grxn1200
        <17->   grxn1728
      }{}

\makeatletter
\newcommand{\normalorbold}{%
  \ifnum\pdf@strcmp{\math@version}{bold}=\z@ bx\else m\fi
}
\makeatother

\begin{document}

\preprint{PITT-PACC-2404}

\title{Lepton-Flavor-Violating ALP Signals with TeV-Scale Muon Beams}

\author{Brian Batell}
\email{batell@pitt.edu}
\affiliation{Pittsburgh Particle Physics, Astrophysics, and Cosmology Center, Department of Physics and Astronomy, University of Pittsburgh, Pittsburgh, PA 15217, USA}

\author{Hooman Davoudiasl}
\email{hooman@bnl.gov}
\affiliation{High Energy Theory Group, Physics Department,Brookhaven National Laboratory, Upton, NY 11973, USA}

\author{Roman Marcarelli}
\email{roman.marcarelli@colorado.edu}
\affiliation{High Energy Theory Group, Physics Department,Brookhaven National Laboratory, Upton, NY 11973, USA}
\affiliation{Department of Physics, University of Colorado, Boulder, Colorado 80309, USA}

\author{Ethan T. Neil}
\email{ethan.neil@colorado.edu}
\affiliation{Department of Physics, University of Colorado, Boulder, Colorado 80309, USA}

\author{Sebastian Trojanowski}
\email{sebastian.trojanowski@ncbj.gov.pl}
\affiliation{National Centre for Nuclear Research, ul.~Pasteura 7, 02-093 Warsaw, Poland}

\begin{abstract}

We explore the feasibility of using TeV-energy muons to probe lepton-flavor-violating (LFV) processes mediated by an axion-like particle (ALP) $a$ with mass $\mathcal{O}(10~\textrm{GeV})$. We focus on $\mu\tau$ LFV interactions and assume that the ALP is coupled to a dark state $\chi$, which can be either less or more massive than $a$. Such a setup is demonstrated to be consistent with $\chi$ being a candidate for dark matter, in the experimentally relevant regime of parameters. We consider the currently operating NA64-$\mu$ experiment and proposed FASER$\nu$2 detector as both the target and the detector for the process $\mu A \to \tau A\, a$, where $A$ is the target nucleus. We also show that a possible future active muon fixed-target experiment operating at a 3 TeV muon collider or in its preparatory phase can provide an impressive reach for the LFV process considered, with future FASER$\nu$2 data providing a pilot study towards that  goal.  The implications of the muon anomalous magnetic moment  $(g-2)_\mu$ measurements for the underlying model, in case of a positive signal, are also examined, and a sample UV completion is outlined.
\end{abstract}

\maketitle
\clearpage

\section{Introduction\label{sec:intro}}

Firm observational evidence for cosmic dark matter (DM) strongly argues for a non-trivial extension of the Standard Model (SM) or perhaps the presence of a new sector of physics.  However, apart from certain theoretical preferences, there is no obvious clue that points to a particular type of DM, or its typical mass scale.  Also, intense experimental effort to find new physics around the weak scale ($\gsim 100$~GeV), a long-time target for DM phenomenology, has so far only resulted in ever strengthening bounds.  The above considerations have led to a broadening of theoretical and experimental DM investigations  to include much lighter candidates, in particular at the GeV scale.  Going to lower mass scales goes hand-in-hand with weaker couplings to  the SM sector, in order to avoid conflict with the existing empirical bounds.  This regime of parameters lends itself well to searches at low energy probes with large effective luminosities, such as fixed target experiments.  

Motivated by the need for a broad view of new-physics models to explain DM and other puzzles of the SM, it is interesting to consider the phenomenology of axion-like particles (ALPs), a generic class of particles that can occur in physics beyond the SM (BSM) in many different contexts~\cite{DiLuzio:2020wdo,Choi:2020rgn}. ALPs occur generically as pseudo-Nambu-Goldstone bosons associated with symmetry breaking, so they can naturally exist at or below the GeV scale even if they are associated with new physics above the weak scale.  ALPs with lepton-flavor-violating (LFV) couplings are particularly interesting, as they can have distinctive phenomenological signatures due to the absence of significant LFV in the SM.  Anomalies related to lepton physics such as the well-known discrepancy between theory and experiment for the value of  the muon $(g-2)_\mu$ further motivate the consideration of leptonic ALP couplings. See Refs.~\cite{Davidson:1981zd, Wilczek:1982rv} for early work on flavor-violating axions.

In this paper, we consider using TeV scale muons produced at high-energy accelerator facilities in order to explore possible $\mu\tau$ LFV interactions of ALPs heavier than the $\tau$ lepton.  For other studies of searches for LFV ALPs at colliders and fixed-target experiments, cf., e.g., Refs~\cite{Bauer:2019gfk,Cornella:2019uxs,Endo:2020mev,Iguro:2020rby,Calibbi:2020jvd,Davoudiasl:2021mjy,Davoudiasl:2021haa,Bauer:2021mvw,Cheung:2021mol,Calibbi:2022izs}. We point out that forward-going muons at the LHC can be used to study such LFV probes using the extension of the FASER$\nu$ detector~\cite{FASER:2019dxq}, the proposed FASER$\nu$2 detector at the envisioned Forward Physics Facility (FPF)~\cite{Feng:2022inv}. This will complement the existing muon fixed-target experimental program in the NA64-$\mu$ experiment at CERN~\cite{Sieber:2021fue}. While those measurements would only go slightly beyond existing ones, they would serve as a testing ground for the general ideas. However, we show that a future muon fixed-target experiment operating at $>\textrm{TeV}$ energies can considerably extend the reach for heavy ALPs with $\mu\tau$ LFV couplings. As discussed in Refs.~\cite{Cesarotti:2022ttv,Cesarotti:2023sje}, such a muon beam-dump experiment could, for example, be related to the preparations of the proposed muon collider with a 3 TeV center-of-mass energy~\cite{Accettura:2023ked,InternationalMuonCollider:2024jyv}. Here, we extend this idea and consider a thin active target detector that could operate in tandem with a beam-dump experiment and use a fraction of the total number of muons on target (MOT) to study the prospects for the ALP search using $1.5~\textrm{TeV}$ muons. In particular, such an experiment can provide a powerful probe of the parameter space relevant to the resolution of the muon $g-2$ anomaly. We demonstrate that it can also probe couplings relevant to DM  physics by providing a simple DM model coupled to the ALP, which in some scenarios can modify the phenomenology through invisible ALP decays. 

This paper is organized as follows. In \cref{sec:ALPscenario}, we introduce the ALP scenario adopted in our work, analyze its predictions for $(g-2)_\mu$ and DM, and discuss other bounds on this model. \Cref{sec:signal} contains a discussion of  experimental prospects for measuring ALP LFV signals in high-energy muon fixed-target experiments. We present our results in \cref{sec:results} and conclude in \cref{sec:conclusions}. In \cref{app:UV}, we present a sample UV completion for which relatively large non-diagonal $\mu\tau$ ALP coupling values can be obtained.

\section{ALP\lowercase{s} with LFV couplings\label{sec:ALPscenario}}

\subsection{Model\label{sec:model}}

The LFV interactions of the ALP, $a$, may be parameterized by the Lagrangian \cite{Cornella:2019uxs}
\begin{align}
{\cal L} & \supset  - i \frac{a}{\Lambda} \sum_{i,j} \bar \ell_i [ (m_j-m_i) v_{ij}  +(m_j+m_i) a_{ij} \gamma^5 ] \ell_j  \nonumber \\
 & \supset  - i a \sum_{i,j}   \bar \ell_i   \frac{C_{ij}}{\Lambda} [ (m_j-m_i) \sin \theta_{ij}     \\
 & +(m_j+m_i) e^{i \phi_{ij}}   \cos \theta_{ij}  \gamma^5 ] \ell_j,   \nonumber
\end{align}
where $C_{ij} = \sqrt{|a_{ij}|^2+|v_{ij}|^2}$, $\theta_{ij} = \arctan |v_{ij}/a_{ij}|$, and $\phi_{ij} = \arg(a_{ij}/v_{ij})$. We take $v_{ij}$ and $a_{ij}$ to be real, implying $\phi_{ij} = 0$. 
Specializing to the case of $\mu\tau$ flavor-violating couplings (and dropping flavor indices on angles), 
\begin{align}
{\cal L}  
 & \supset  - i   \frac{C_{\mu \tau}}{\Lambda}  a
  \bar \mu \, [ (m_\tau-m_\mu) \sin \theta  +(m_\tau+m_\mu)    \cos \theta  \, \gamma^5 ] \, \tau \nonumber \\
&  + {\rm h.c.} \label{eq:L_mutau_simple}
\end{align}
Taking the limit $m_\mu \rightarrow 0$ and defining $g_{\mu\tau} \equiv C_{\mu\tau} m_\tau/\Lambda$, we have 
\begin{align}
{\cal L}  
 & \supset  - i  g_{\mu\tau}  a
  \bar \mu \,  [ \sin \theta  +    \cos \theta \,  \gamma^5 ]  \,\tau   + {\rm h.c.}
\end{align}
To give a rough sense of the size of the coupling, for $C_{\mu\tau}/\Lambda = (1\,{\rm  TeV})^{-1}$, we obtain $g_{\mu \tau } \simeq 2 \times 10^{-3}$. 
However, it is possible in certain UV completions that couplings as large as $g_{\mu \tau } \sim {\cal O}(10^{-1})$ can be obtained. We present an explicit UV model realizing this possibility in   \cref{app:UV}. 

In addition to the $C_{\mu \tau}$ coupling, we include a coupling to a hidden Dirac fermion $\chi$ which will serve as a DM  candidate.  The interaction Lagrangian is 
\begin{align}
{\cal L} \supset - i \frac{C_\chi}{\Lambda_\chi}\, 2m_\chi  a \bar \chi \gamma^5 \chi  = - i g_\chi a \bar \chi \gamma^5 \chi,
\end{align}
where we have defined $g_\chi \equiv 2\,C_\chi m_\chi / \Lambda_\chi$. Note that for the ALP coupling to DM we can in principle consider a much lower UV scale $\Lambda_\chi$ than in the case of the LFV coupling discussed above, such that $g_\chi$ could perhaps be as large as order unity. 

We have in mind two scenarios, which will lead to two distinct experimental signatures:
\begin{itemize}
\item {\it Visibly decaying ALP:} This scenario involves an ALP that decays via $a\rightarrow \mu^\pm \tau^\mp$. This will occur when $m_a > m_\mu + m_\tau$ and when the ALP decays to DM are kinematically forbidden, $m_a < 2 m_\chi$. The rate for this decay is
\begin{equation}
    \Gamma(a \rightarrow \mu^{\pm}\tau^{\mp}) \approx \frac{C_{\mu \tau}^2m_\tau^2}{8\pi \Lambda^2}m_a(1 - m_\tau^2/m_a^2)^2.
\end{equation}
This corresponds to a characteristic decay length for $m_a\gg m_\tau$ of order
\begin{equation}
c\,\tau_{a}\simeq (10^{-8}~\textrm{cm})\,\left(\frac{C_{\mu\tau}/\Lambda}{\textrm{TeV}^{-1}}\right)^{-2}\,\left(\frac{15~\textrm{GeV}}{m_a}\right).
\end{equation}
In this case, even boosted ALPs with a few hundred GeV energy decay promptly and do not leave displaced signatures in the detector. The only exceptions are scenarios with $m_a\sim m_\tau+m_\mu$ and low values of the coupling constant, e.g., for $C_{\mu\tau}/\Lambda \sim 10^{-2}~\textrm{TeV}^{-1}$ and $m_a = 2~\textrm{GeV}$, we find $\gamma\beta c\tau_a\sim \textrm{a few}~\textrm{cm}$ for $E_a\sim \textrm{several hundred GeV}$. This could lead to displaced decays of the ALP in the detector and further help identify BSM events. However, as we will see in \cref{sec:signalBG,sec:results}, this only affects a very small region of the parameter space relevant to the sensitivity reach of a future proposed muon fixed-target experiment. Therefore, we assume that ALPs decay promptly in the analysis below.

\item {\it Invisibly decaying ALP:}   This scenario involves an ALP that decays to DM, leading to missing energy. Such decays will typically dominate if $m_a > 2 m_\chi$, given the expected larger couplings of ALPs to the dark sector. Note that in this scenario there are strong constraints from the LFV decay $\tau \rightarrow \mu a$ for ALP masses below $m_a < m_\tau - m_\mu$~\cite{Iguro:2020rby}. Instead, we focus on heavier ALPs in the following. The rate for the invisible ALP decay is
\begin{equation}
    \Gamma(a\rightarrow \bar{\chi}\chi) = \frac{C_{\chi}^2m_\chi^2}{2\pi\Lambda_\chi^2}m_a\sqrt{1 - 4m_\chi^2/m_a^2}\,.
\end{equation}

\end{itemize}

\subsection{Muon anomalous magnetic moment\label{sec:gm2}}

The ALP LFV coupling generates a contribution to  $a_\mu \equiv (g-2)_\mu/2$ at one loop. Considering the regime
$m_\tau \gg m_\mu$, the result is 
 \begin{equation}
 a_\mu^a \simeq -  \frac{g_{\mu\tau}^2 }{16 \pi^2} \frac{m_\mu^2}{m_\tau^2}\left[f(x_\tau)  + \frac{m_\tau}{m_\mu} g (x_\tau) \cos 2\theta \right],
 \end{equation}
where $x = m_a^2 / m_\tau^2$ and
\begin{align}
f(x) & = \frac{2 x^2 (2x-1) \log x}{(x-1)^4} -\frac{5-19x+20x^2}{3(x-1)^3},\nonumber \\
g(x) & =  \frac{2 x^2 \log x}{(x-1)^3} + \frac{1-3x}{(x-1)^2}.
\end{align}
Furthermore, for the sake of completeness, in the low ALP mass regime the expression simplifies to (assuming $|\cos 2\theta| \gg m_\mu/m_\tau)$
\begin{equation}
 a_\mu^a \simeq -  \frac{g_{\mu\tau}^2 }{16 \pi^2} \frac{m_\mu}{m_\tau} \cos 2\theta, 
 \label{eq:amu-approx}
 \end{equation}
though this regime is typically outside of the parameter region of interest in our study. 

The current  experimental world average~\cite{Muong-2:2023cdq} and SM theory prediction~\cite{Aoyama:2020ynm} are 
\begin{align}
a_{\mu}^{\rm Exp} & =  116\,592\,059 (22) \times 10^{-11}, \nonumber \\
a_{\mu}^{\rm SM} & =  116\,591\,810(43)  \times 10^{-11}.
\end{align}
This implies a discrepancy of 
\begin{align}
a_{\mu}^{\rm Exp} - a_{\mu}^{\rm SM}  & =  (249 \pm 48) \times 10^{-11},
\label{eq:delta-amu}
\end{align}
or about 5.1 $\sigma$; however, there is some tension between data-driven dispersive and lattice QCD inputs (particularly from Ref.~\cite{Borsanyi:2020mff}) to the SM theory number; see Ref.~\cite{Colangelo:2022jxc} for an updated overview of the situation.
We see from Eq.~(\ref{eq:amu-approx}) that for  $\cos 2\theta  < 0$, or $ \pi/4 < \theta < 3 \pi/4 $ we obtain the positive shift in Eq.~(\ref{eq:delta-amu}). For $\cos 2\theta  < 0$ and order unity, we thus require an effective LFV coupling $g_{\mu\tau} \sim  {\rm few} \times 10^{-3}$ or larger to explain the discrepancy.  

\subsection{Dark matter\label{sec:DM}}

For the Dirac fermion DM  candidate $\chi$ considered here, there are two potential annihilation channels that may be relevant for setting the relic abundance. First, for $2m_\chi > m_\tau + m_\mu$, the DM can annihilate to the SM via $\chi \bar \chi \rightarrow \mu^- \tau^+$ and $\chi \bar \chi \rightarrow \tau^- \mu^+$ via  $s$-channel ALP exchange. We refer to the sum of these channels simply as $\chi \bar \chi \rightarrow \mu \tau$. This channel typically dominates when the DM cannot kinematically annihilate to a pair of on-shell ALPs. The thermally averaged annihilation cross section in the nonrelativistic regime is 
\begin{equation}
\langle \sigma  v \rangle   \simeq \frac{g_{\mu\tau}^2  g_\chi^2}{ 16 \pi } \frac{(4 m_\chi^2 - m_\tau^2)^2}{m_\chi^2 (4 m_\chi^2-m_a^2)^2}.
\end{equation}
For Dirac $\chi$, the observed DM relic abundance is obtained for an annihilation cross section of $\langle \sigma v \rangle  \simeq 4.4 \times 10^{-26}$ cm$^3$ s$^{-1}$~\cite{Steigman:2012nb}. Note, however, that the annihilation in this case occurs in the $s$-wave and may thus be strongly constrained by DM indirect detection (ID) searches and the CMB if $m_\chi \lesssim 10$ GeV. To have a viable scenario for lighter DM candidates, we can consider asymmetric DM, which would require large annihilation cross sections in order to deplete the symmetric component.  

\begin{figure}
 \centering
    \includegraphics[width=1\linewidth]{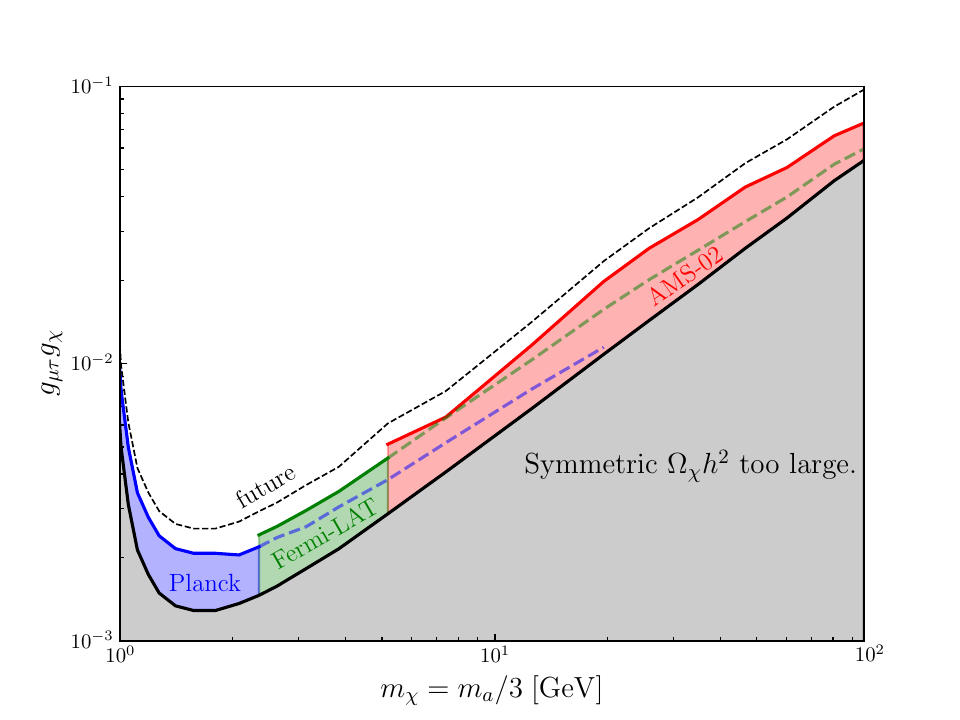}
    \caption{Bounds on the asymmetric DM scenario, in which the DM species efficiently annihilate into $\mu\tau$ final states via the ALP LFV mediator. We assume $m_a = 3\,m_{\chi}$ in the plot. The constraints on the product of the dark and SM coupling constants are shown, $g_{\mu\tau} g_\chi$. The gray-shaded region at the bottom corresponds to the overproduction of a symmetric DM component in the early Universe. The colorful shaded regions above present current indirect detection bounds from Planck~\cite{Planck:2018vyg}, Fermi-LAT~\cite{McDaniel:2023bju}, and AMS-02~\cite{John:2021ugy}. The future sensitivity of such searches is schematically illustrated with a black dashed line following Ref.~\cite{Cooley:2022ufh}.}
    \label{fig:XX-mutau}
\end{figure}

\Cref{fig:XX-mutau} shows parameter values in the $m_\chi - g_{\mu\tau}g_\chi$ plane where the relic density of the symmetric DM component can be suppressed below the measured total DM relic abundance. For a given DM mass and fixed value of the dark coupling constant $g_\chi$, this sets the lower bound on the $g_{\mu\tau}$ coupling constant. We have fixed $m_a = 3 m_\chi$ in the plot. Even a subdominant symmetric DM component can be subject to DM ID constraints~\cite{Graesser:2011wi}. We illustrate this by considering bounds from the Planck CMB observations~\cite{Planck:2018vyg}, Fermi-LAT $\gamma$-ray flux limits from dwarf spheroidal galaxies~\cite{McDaniel:2023bju}, and positron measurements by AMS-02~\cite{John:2021ugy}. These bounds are typically reported for flavor-diagonal DM annihilation final states, i.e., $\mu^+\mu^-$ or $\tau^+\tau^-$. In the plot, for illustration, we present the impact of the parameter space of the stronger of such bounds from each of the experiments above. A more precise modeling should consider a specific $\mu\tau$ annihilation final state relevant to our scenario that could introduce mild corrections to the bound on $g_{\mu\tau}$. We also schematically show future prospects for DM ID searches. This assumes an order of magnitude improvement in the cross section limit with respect to the current constraints~\cite{Cooley:2022ufh} and takes into account the change of the present fractional asymmetry between the DM species with the variation in $\langle\sigma v\rangle$. 

If instead the ALP is relatively light compared to the DM, then annihilation via $\chi \bar \chi \rightarrow a a$ can set the DM abundance. The annihilation cross section is \cite{Armando:2023zwz}
\begin{align}
\langle \sigma  v \rangle   \simeq  \frac{6}{x_{\rm f.o.}}  \frac{ g_\chi^4}{ 24 \pi } \frac{ m_\chi^2 (m_\chi^2 - m_a^2)^2}{ (2 m_\chi^2-m_a^2)^4} \left( 1- \frac{m_a^2}{m_\chi^2}  \right)^{1/2}.
\end{align}
In this case the cross section does not depend on the mediator coupling to the SM, and thus there is no firm relic density target. For a given DM and ALP mass, one can obtain the correct relic abundance by adjusting the ALP - DM coupling $g_\chi$. Nevertheless, this scenario provides a motivation for our visible ALP signal at accelerator experiments. Note also that the annihilation proceeds in the $p$-wave, thus evading the otherwise stringent constraints from the CMB and ID. 

\subsection{Other bounds\label{sec:otherbounds}} 

While we focus on the phenomenology of the ALP model with only one non-vanishing ALP-SM interaction, specifically the LFV $g_{\mu\tau}$ coupling, it is important to note that in realistic UV completions other ALP couplings can also arise. In \cref{app:UV}, we provide an explicit example of such a model, which allows for large values of $g_{\mu\tau}$ and generates additional ALP-SM interactions, and discuss further bounds related to this scenario. At a more general level, the presence of additional ALP couplings to the charged-lepton sector of the SM could introduce stringent bounds on our scenario, and we comment on this below.

\textbf{Other ALP couplings:} As discussed above, important bounds on our scenario are related to the predicted BSM contribution to the anomalous magnetic moment of the muon. Although we focus only on the case where the off-diagonal coupling $g_{\mu \tau}$ is present, the lepton-flavor-diagonal pseudoscalar coupling $g_{\mu \mu}$ generically gives a negative contribution to $a_{\mu}$, see \cite{Marciano:2016yhf,Bauer:2019gfk,Cornella:2019uxs}, which would allow even larger values of $g_{\mu \tau}$ to be consistent with the observed discrepancy. However, the presence of a significant $g_{\mu \mu}$ would lead to strong bounds from $\tau \rightarrow \mu \gamma$ and $\tau \rightarrow 3\mu$ \cite{Cornella:2019uxs,Bauer:2021mvw} that would exclude such large $g_{\mu \tau}$ couplings. A similar conclusion holds if an additional $g_{\tau\tau}$ coupling is present, while a non-zero $g_{ee}$ would induce further bounds from  $\tau\to\mu ee$ decays. 
If such flavor-diagonal ALP couplings are not present at tree level, they could still be radiatively induced if, for example, there are non-zero ALP-gauge boson couplings~\cite{Bauer:2017ris}. However, in this case the generated coupling is of order $\alpha^2$ and thus remains negligible compared to the larger values of $g_{\mu \tau}$ considered here.
It should be noted that if only $g_{\mu\tau}$ is present, diagonal couplings $g_{\mu\mu}$ and  $g_{\tau\tau}$ are not radiatively generated due to the presence of two $Z_2$ symmetries: $Z_{2,a\mu}$ ($Z_{2,a\tau}$) under which both $a$ and $\mu$ ($\tau$) are odd while all other fields are even. 

We also note that if additional off-diagonal couplings between ALPs and leptons are present, i.e., $g_{e \mu}$ or $g_{e \tau}$, then stringent bounds from $\tau\to e\gamma$ and $\mu\to e\gamma$ constrain the available parameter space of the model.
These couplings will not be significantly generated via radiative processes from only $g_{\mu \tau}$ since this coupling preserves $L_\mu + L_\tau$ (along with the two $Z_2$ symmetries mentioned above). We neglect the ALP couplings other than $g_{\mu\tau}$ for the remainder of this work.

\textbf{Bounds from \boldmath $\tau$ decay:} The $g_{\mu\tau}$ coupling alone induces other BSM processes that could be probed. In particular, we focus below on relatively heavy ALPs with the mass $m_a>m_\tau-m_\mu$, such that stringent bounds~\cite{ARGUS:1995bjh,Calibbi:2020jvd} on $2$-body tau decays into the muon and missing energy, $\tau \to \mu +\textrm{inv.}$, do not apply. We note that these bounds do not automatically constrain the model with the invisibly decaying ALP, i.e., the process $\tau \to \mu (a^\ast\to \chi\bar{\chi})$, as they rely on the expected peak in the final-state muon energy in the (pseudo) rest frame of the tau lepton, which is characteristic for a $2$-body decay kinematics. 

Instead, the $3$-body decay kinematics resembles that of the SM decay process, $\tau \to \mu\nu\nu$. Bounds on this scenario can be derived based on the ratio between the branching fractions of the $\tau$ leptonic decays into muons and electrons. For light DM and the ALP mass $m_a\sim 2~\textrm{GeV}$, the bounds could be as severe as $g_{\mu\tau}g_\chi\lesssim 10^{-6}$, cf. Ref.~\cite{Altmannshofer:2016brv} for related discussion for an intermediate vector boson $Z^\prime$. In the following, we will assume the mass ratio between the ALP and DM to be  $m_a/m_\chi = 3$. The considered bound could then constrain the parameter space of the invisibly decaying ALP scenario for $m_a\lesssim 2.5~\textrm{GeV}$. However, as the ALP mass approaches this upper limit, one expects that the bounds become less severe, given that kinematics of the tau lepton decay into massive DM species $\chi$ would not resemble the decay into neutrinos. We note that a precise bound could be obtained with a dedicated experimental analysis. 

\textbf{Bounds on \boldmath ${a\to\chi\bar{\chi}}$ from ${h}$ and ${Z}$ decays:} Exotic decays of the Higgs boson can also lead to significant bounds on the off-diagonal coupling $g_{\mu \tau}$. In particular, LFV decays $h\to \mu\tau a$ are expected in this case, where the ALP can subsequently decay invisibly.  In this case, existing searches for $h\to\mu\tau$ decays could apply. However, we note that the current bounds are based on the approximate reconstruction of the invariant mass of the $\mu-\tau$ system~\cite{ATLAS:2019pmk,CMS:2021rsq,Barman:2022iwj}. A priori, this is not applicable to the $3$-body kinematics relevant to our scenario, and a dedicated analysis should be performed. 

Further constraints come from $Z$ boson decays. The predicted $3$-body decay branching fraction of the $Z$ boson to the ALP and a $\mu\tau$ pair is estimated to be $\mathcal{B}(Z\to \mu^\pm\tau^\mp a)\sim 2\times 10^{-9}\,[(C_{\mu\tau}/\Lambda)/\textrm{TeV}^{-1}]^2$, using the results of  Ref.~\cite{Calibbi:2022izs} for the lepton-flavor-conserving case. See also Ref.~\cite{Altmannshofer:2022ckw} for recent studies of $W$ boson decays to ALPs. This can be compared with the bound on the LFV decay branching fraction derived by the ATLAS experiment, $\mathcal{B}(Z\to\mu\tau)< 6.5\times 10^{-6}$~\cite{ATLAS:2020zlz,ATLAS:2021bdj}. We  translate the preceding bound  into the approximate constraint on the ALP coupling constant for the invisibly decaying ALP scenario of order $C_{\mu\tau}/\Lambda \lesssim 60~\textrm{TeV}^{-1}$.  Note, however, that the $3$-body kinematics of the $Z$-boson decay into the ALP could affect the bound, especially for  larger ALP masses.

\textbf{Bounds on \boldmath  ${a\to\mu\tau}$ from $h$ and $Z$ decays:} Visibly decaying ALPs could be constrained based on searches for the Higgs boson decaying into two muons and two tau leptons. Existing bounds, obtained under the assumption of lepton flavor conservation, assume a cascade decay process with intermediate ALPs, $h\to aa\to\mu\mu\tau\tau$ and the reconstruction relies, i.a., on measuring the invariant mass of the di-muon pair, cf., e.g., Refs~\cite{ATLAS:2015unc,CMS:2020ffa}. However, the initial $3$-body decay kinematics followed by a $2$-body ALP decay, which is characteristic of the model considered here, differs from these scenarios and will require other signal selection and background mitigation strategies. Similar discussion applies to past bounds based on rare decays of $B$ mesons~\cite{Cornella:2019uxs,LHCb:2019ujz}.

The visibly decaying ALP could also be constrained in measurements of the $Z\to2\mu 2\tau$ decays. The CMS collaboration has recently constrained the ratio $\mathcal{R}_{\tau\tau\mu\mu}$ between the branching fraction of this decay and the one for $Z\to 4\mu$ process to be less than $6.9$ times the SM expectation, which was $\mathcal{R}_{\tau\tau\mu\mu}\simeq 0.9$ with the cuts used in the analysis~\cite{CMS:2024uqa}. We note that the branching fraction into four muons is measured to be $\mathcal{B}(Z\to 4\mu)\simeq (1.20 \pm 0.08) \times 10^{-6}$ assuming $m_{\mu^+\mu^-}>4~\textrm{GeV}$ for all pairs of oppositely-charged muons~\cite{CMS:2017dzg,CMS:2018vcg}. Given the aforementioned predicted BSM branching fraction of $Z\to a\mu\tau$, we find that too large BSM contributions to the $Z\to2\mu 2\tau$ decays for the visibly decaying ALP are excluded, which leads to the bound on the coupling constant $C_{\mu\tau}/\Lambda \lesssim 60~\textrm{TeV}^{-1}$.

\textbf{Explicit LFV ALP signature at colliders:} While the above bounds on visibly decaying ALPs were based on lepton-flavor-conserving signatures, they can also be constrained in searches for explicit LFV at lepton colliders~\cite{Dev:2017ftk} and in Higgs decays at the LHC~\cite{Davoudiasl:2021haa}. This relies on the presence of same-sign muons and tau leptons in the final state, e.g., $e^+e^-\to \gamma^{(\ast)},Z^{(\ast)}\to \mu^+\tau^-(a^{(\ast)}\to\mu^+\tau^-)$. Especially relevant are future colliders operating at $Z$ pole, e.g., FCC-ee and CEPC, which will constrain the ALP coupling strength to $C_{\mu\tau}\lesssim \mathcal{O}(1~\textrm{TeV}^{-1})$ or $g_{\mu\tau}\lesssim \textrm{a few}\times 10^{-3}$ for the ALP mass up to tens of GeV~\cite{Dev:2017ftk}. Similar constraints are expected from searches in the proposed $\mu$TRISTAN $\mu^+\mu^+$ and $\mu^+e^-$ colliders~\cite{Calibbi:2024rcm}. Even stronger bounds can be obtained for Belle-II, down to $g_{\mu\tau}\lesssim 10^{-4}$ for $m_a\sim 2~\textrm{GeV}$~\cite{Iguro:2020rby}. However, these bounds quickly diminish with increasing ALP mass and are only relevant for $m_a\lesssim 10~\textrm{GeV}$. We conclude that future collider-based searches for explicit LFV in ALP-induced processes will provide an independent test of the region favored by the $(g-2)_\mu$ anomaly in our model, which is complementary to the sensitivity of fixed-target experiments discussed below. Finally, we also note that it would also be worthwhile to explore LFV signals of ALPs in Higgs and $Z$ decays at the LHC for our $\mu\tau$ LFV ALP scenario. 

\section{LFV Signal\label{sec:signal}}

One way that LFV ALPs can be produced in an accelerator experiment is via electromagnetic interactions between a charged lepton and a heavy nucleus. This has been considered in the past for electron-nucleus interactions at the EIC \cite{Davoudiasl:2021mjy}, which allows one to constrain the $e\tau$ coupling of the ALP. Producing GeV-scale ALPs with an $e\tau$ coupling in this way requires a TeV electron in the rest frame of the ion, which is possible at the EIC due to the high energy of the ion in the lab frame. By analogy, to constrain the $\mu\tau$ coupling of the model to a similar degree would require TeV-scale muons incident on nuclei.

The $\mu\to\tau$ conversion process with a sizable associated missing momentum was previously studied in the context of searches in muon fixed-target experiments in both the deep-inelastic scattering (DIS) regime~\cite{Gninenko:2001id,Gninenko:2018num} and due to the coherent production of GeV-scale BSM scalar particles in muon interactions with nuclei~\cite{Radics:2023tkn}; cf. also Ref~\cite{Black:2002wh} for past experimental bounds on effective LFV $\mu\tau$ operators with additional couplings to quarks. Further production and detection modes appear if new physics species produced in muon scatterings are long-lived and can visibly decay in a distant detector~\cite{Ema:2022afm}. 

In the following, we focus on probing a prompt-decay regime of ALPs with masses above the tau lepton mass up to tens of GeV using ongoing and future high-energy muon fixed-target experiments. Producing such a massive ALP in a muon fixed-target experiment requires employing a high-energy muon beam. Such energetic muons are available in the ongoing NA64-$\mu$ experiment at CERN. Even higher energy muons are abundantly produced in the forward kinematic region of the LHC and can be measured in the proposed FPF. However, as we will see, in both these cases the limited available muon flux leads to observable event rates only for very large ALP couplings. A much more powerful probe of the $\mu \tau$ coupling parameter space would be available at a beam-dump experiment associated with a future multi-TeV muon collider, which we also explore. 

\subsection{ALP production}

\begin{figure}[t]
    \centering
    \includegraphics[width=0.75\linewidth]{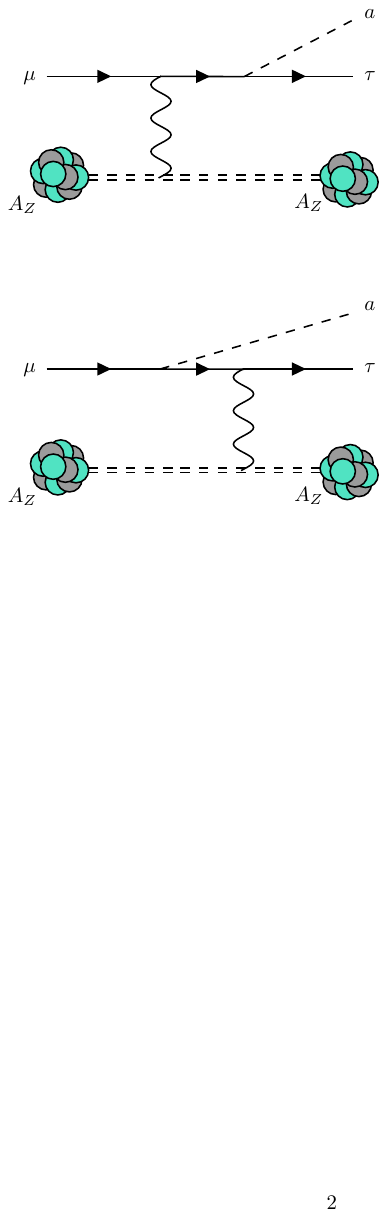}
    \caption{The ALP production process in a coherent muon scattering off a nucleus. 
    }
    \label{fig:diagrams}
\end{figure}

In our analysis, we consider ALPs produced from a high-energy muon electromagnetically interacting with atoms within a material. The Improved Weizsacker-Williams approximation \cite{Kim:1973he} is often employed to simplify such calculations, but it has been shown to not always be accurate in the large-mass regime of the produced particle \cite{Liu:2016mqv,Liu:2017htz}. Instead, we integrate the full cross-section of the $2 \rightarrow 3$ process $\mu N \rightarrow \tau N a$, then use the narrow width approximation for decay of the ALP. Along with a more accurate cross-section, this approach allows us to analyze the energy and angular distributions of the final-state ALP.  Details of the cross-section integration can be found in the appendix of Ref.~\cite{Davoudiasl:2021mjy}. 

One might expect that a large momentum-transfer is necessary to create GeV-scale ALPs, but this is not always the case, as the required momentum transfer is dependent on both the ALP mass and the incident muon energy. In particular, an ALP of mass $m_a$ can be produced from a muon of energy $E_\mu$ exchanging a photon with a nucleus provided the momentum transfer  is larger than $t_{\rm min} \sim (m_a^2 / 2E_\mu)^2$. Since the coherent scattering cross section is increasing for low momentum transfer $t$, it grows with the incident muon energy. Additional cross-section could be gained by considering photon exchange in the deep-inelastic scattering (DIS) regime, but this is especially relevant when $t_{\rm min} \gsim 1~{\rm GeV}$, which corresponds to $m_a \gsim 40~{\rm GeV}$ for a $1~{\rm TeV}$ muon beam. Here we only consider coherent and diffractive interactions between the muon and the nucleons of the nucleus for which backgrounds can be rejected more easily, and we impose a cut-off on $t$ of $1~{\rm GeV}$ rather than considering DIS effects. 

For coherent interactions, the cross-section is enhanced by a factor of $Z^2$, and for diffractive interactions, it is still enhanced by a factor of around $Z$ (although for larger ALP masses, this effect is diminished by a smaller form factor). The diagram for ALP production is shown in \cref{fig:diagrams}. For simplicity, we treat the nucleus as a scalar with mass $M$ and charge $Ze$, which is a reasonable approximation in the absence of spin-related effects. The photon-ion interaction vertex is then given by
\begin{align}
    iV^\mu = ieZ\hat{F}(q^2)(P_1^\mu + P_2^\mu), \label{eq:vertex}
\end{align}
where $\hat{F}(q^2)$ is a normalized form factor that incorporates both the coherent and diffractive interaction of the nucleus. In particular, we decompose the squared form factor into an elastic and inelastic component, 
\begin{align}
    Z^2\hat{F}(q^2)^2 &= G_{\rm el}^{\rm atom}(q^2)G_{\rm el}^{\rm nuc}(q^2) \nonumber \\
    \ \ &+ G_{\rm inel}^{\rm atom}(q^2)G_{\rm inel}^{\rm nuc}(q^2),
\end{align}
where the superscripts ``atom'' and ``nuc'' refer to the atomic and nuclear contributions, and the subscripts ``el'' and ``inel'' correspond to elastic and inelastic form factors. For the atomic form factors, we take \cite{Tsai:1973py} 
\begin{align}
    G_{\rm el}^{\rm atom}(t) &= \left[\frac{a^2t}{1+a^2t}\right]^2,\\  G_{\rm inel}^{\rm atom}(t) &= \left[\frac{a'^2t}{1+a'^2t}\right]^2,
\end{align}
with $a = 111Z^{-1/3}/m_e$ and $a' = 571.4Z^{-2/3}/m_e$.

For the nuclear inelastic form factor, we use the dipole approximation for the electromagnetic form factors of the nucleons within the nucleus \cite{Tsai:1973py},
\begin{align}
    G_{\rm inel}^{\rm nuc}(t) = &\frac{1}{(1 + t/t_0)^4 (1 + t/(4m_p^2))}\left[Z(1+(\mu_p^2/4m_p^2)t)\right. \nonumber \\ 
    &\left. + (A-Z)(\mu_n^2/4m_p^2)t\right],
\end{align}
with $t_0 = 0.71~{\rm GeV}^2$, $\mu_p = 2.79$, and $\mu_n = 1.91$. The first term represents the elastic form factor of the proton, and the second term represents the elastic form factor of the neutron. The neutron's contribution is only comparable to the proton's when $t \sim m_p^2$, as expected. The dipole approximation becomes less accurate at higher momentum transfer,\cite{Puckett:2017flj} so we impose a cut-off on the form factors at $t = 1~{\rm GeV}$.

\begin{figure}
    \centering
\includegraphics[width = \linewidth]{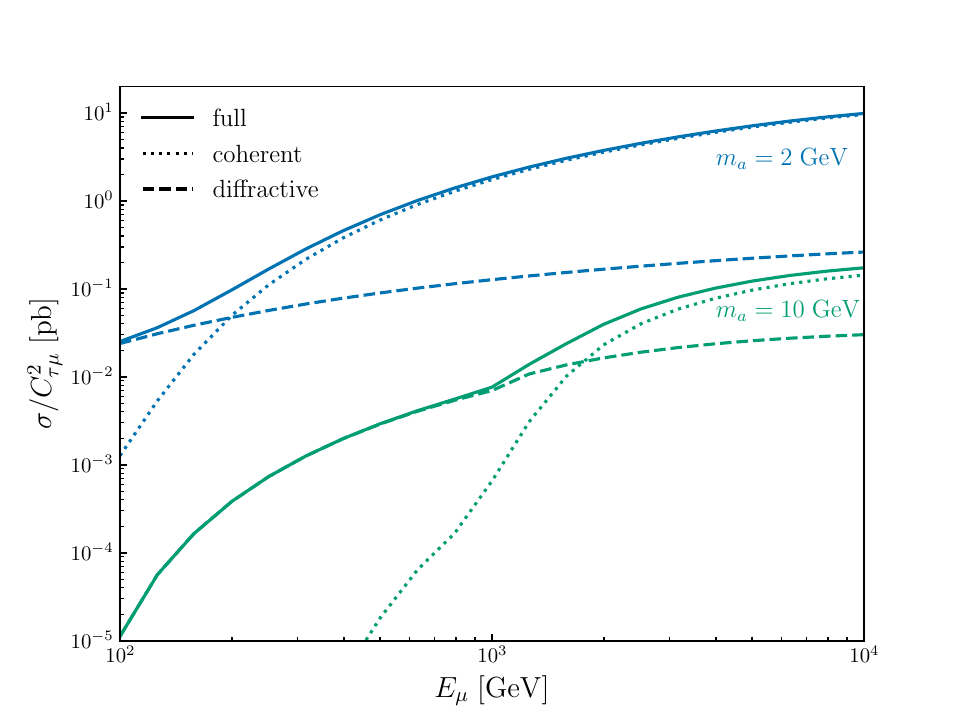}
    \caption{Diffractive (dashed), coherent (dotted), and total (solid) cross-section vs. incident muon energy for $m_a = 2~{\rm GeV}$ (blue) and $m_a=10~{\rm GeV}$ (green). Here we have fixed $\Lambda = 1$ TeV.
    }
    \label{fig:coh_inc_energy}
\end{figure}

Finally, for the nuclear elastic form factor, we use an approximation of the Fourier-transform of the Woods-Saxon distribution \cite{Klein:1999qj}, given by 
\begin{align}
    &G_{\rm el}^{\rm nuc}(t) = \frac{9Z^2}{t^3 R_A^6} \times \nonumber \\
    &\left[\left(\sin{(R_A\sqrt{t})}-R_A\sqrt{t}\cos{(R_A\sqrt{t})}\right) \frac{1}{1+a_0^2 t}\right]^2,
\end{align}
where $a_0 = 0.79\,{\rm fm}$ and $R_A = (1.2\,{\rm fm})A^{1/3}$. Note that for small $t$, this agrees with the form of the coherent nuclear form factor used in Ref.~\cite{Tsai:1973py}, but falls off more quickly at large $t$ ($\sim 1/t^5$ instead of $\sim 1/t^2$). The contribution of the coherent and diffractive form factors to the total cross-section for ALP masses $m_a = {2~\rm GeV}$ and $m_a = {10~\rm GeV}$ is shown in \cref{fig:coh_inc_energy}.

With the above  form factor, one can use the vertex \cref{eq:vertex} to compute the total production cross-section. We refer the reader to Refs.~\cite{Liu:2016mqv,Liu:2017htz,Davoudiasl:2021mjy} for details and formulae regarding the full $2 \rightarrow 3$ cross section.

\subsection{Experiments\label{sec:experiments}}

We consider two distinct scenarios for the decay of the ALP: visible decays into $\mu^\pm\tau^\mp$, and invisible decays into the dark fermions $\bar{\chi} \chi$. Different search strategies must be designed to target these distinct ALP decay modes. The discovery prospects also depend on the capability to reconstruct short tau lepton tracks and determine the incident muon charge. Below we first briefly describe the experimental facilities of interest to our work and then discuss sample signal selection strategies for both of the considered ALP decay modes. 

\medskip
\textbf{The NA64-$\mathbf{\mu}$ experiment at CERN} employs 
the
high-intensity SPS muon beam with $E_\mu \simeq 160~\textrm{GeV}$~\cite{Sieber:2021fue}. The muons are dumped on lead-scintillator (Pb-Sc) target material serving as an electromagnetic calorimeter (ECAL) with approximately forty radiation lengths of lead~\cite{Gninenko:2640930}. Below we present the projected exclusion bounds assuming $10^{13}$ MOT. We also estimate the current bounds based on $2\times 10^{10}$ MOT~\cite{Andreev:2024sgn}.

Other specific muon beam-dump experiments have also been proposed, such as the Muon Missing Momentum ($\textrm{M}^3$) experiment at Fermilab~\cite{Kahn:2018cqs}. This experiment would, however, employ a muon beam with a lower energy of $E_\mu = 15~\textrm{GeV}$, resulting in a substantially smaller production cross section for heavy ALPs. Instead, the same high-energy CERN muon beam with $E_{\mu} \sim (150-160)~\textrm{GeV}$ could also be used to search for light BSM particles in the MUonE experiment~\cite{GrillidiCortona:2022kbq}. In this case, however, the low-$Z$  target materials~\cite{Abbiendi:2677471} result in a significantly suppressed coherent ALP production cross section compared with the NA64-$\mu$ detector setup.

\medskip
\textbf{The proposed Forward Physics Facility at the LHC} will offer an opportunity to exploit even more energetic muons, up to ${\cal O}(\rm TeV)$, which are created in the decays of hadrons produced in $pp$ collisions at the ATLAS interaction point (IP). Despite the impact of the LHC magnets and shielding by the infrastructure, a significant number of both negative and positive muons can reach the far-forward experiments, which are proposed to be placed about $620~\textrm{m}$ away from the IP. Our sensitivity studies employ the muon flux and spectrum predictions obtained for the FPF by the CERN FLUKA (STI) team~\cite{sabate-gilarte:ipac2023-mopl018}. It extends up to $E_\mu\sim \textrm{a few TeV}$ energy. In the following, we assume that the FPF experiments cannot determine the incident high-energy muon charge ($\mu^+$ or $\mu^-$), while the outgoing muon charge can be measured.

In our analysis, we specifically focus on the proposed FASER$\nu$2 experiment~\cite{Anchordoqui:2021ghd,Feng:2022inv}, which will employ $20~\textrm{tons}$ of tungsten target material and emulsion detector technology. This detection technique is currently used, i.a., in the FASER$\nu$~\cite{FASER:2019dxq,FASER:2020gpr} and SND@LHC~\cite{SHiP:2020sos,SNDLHC:2022ihg} detectors at the LHC. The total length of the tungsten layers in FASER$\nu$2 is $6.4~\textrm{m}$. Emulsion detectors feature excellent spatial resolution, allowing for a detailed study of high energy neutrino interactions, including the precise measurement of tau lepton tracks produced in charge current  interactions of $\nu_\tau$s. This can be done with a resolution of $\delta_{\textrm{FASER$\nu$2}} = 2~\textrm{mm}$, which corresponds to the thickness of the tungsten layers interleaved with emulsion. In the following, we will rely on these expected FASER$\nu$2 capabilities to study muon-induced ALP signals. 

We note that the central LHC detectors have also been proposed to work in the muon beam-dump mode and search for the missing-momentum signature~\cite{Galon:2019owl}. In this case, due to the required purity of the muon sample, only muons from $W$ and $Z$ boson decays were considered with an average energy of order tens of GeV. For this reason, and since the dominant source of high-energy muons at the LHC is due to decays of hadrons with lower masses and $p_T$, we focus on forward-going muons and the proposed FASER$\nu$2 detector in the following. 

\medskip
\textbf{A high-energy muon fixed-target experiment} employing future TeV-scale muon beams will be able to harness both advantages -- high intensities and high energies -- of the aforementioned experiments at CERN. Recently there has been renewed interest in the possibility of a multi-TeV muon collider~\cite{Delahaye:2019omf,Delahaye:2013jla}, which would offer a new window into the energy frontier. In this context, it is interesting to explore other physics applications of high energy muon beams. 

Here we consider the possibility of a high energy muon fixed target experiment, as has been recently investigated in Refs.~\cite{Cesarotti:2022ttv,Cesarotti:2023sje}. Such an experiment could be constructed as one component of the muon collider facility, or developed even earlier during the muon beam R$\&$D phase. The original proposal considered a TeV-energy muon beam dumped on a thick target ($\sim 10$m) along with a detector located $\sim 100$ m downstream. 
Such a setup is well suited to searches for light, weakly-coupled, long lived particles. In order to facilitate such new physics searches, a strong magnet should be placed downstream in order to deflect the muons leaving the target material and mitigate backgrounds. In contrast, new light BSM particles will not be affected by the magnet and can travel forward to the decay volume located further downstream and preceded by a front veto. The visible decay products of the BSM species can then be analyzed in the main detector, which should be capable of tracking and measuring the energy of charged particles, including muons. The predicted BSM sensitivity depends on the total number of MOT, which can be as large as $10^{20}$ for the multi-year run.

\begin{figure}[t]
    \centering
    \includegraphics[width=\linewidth]{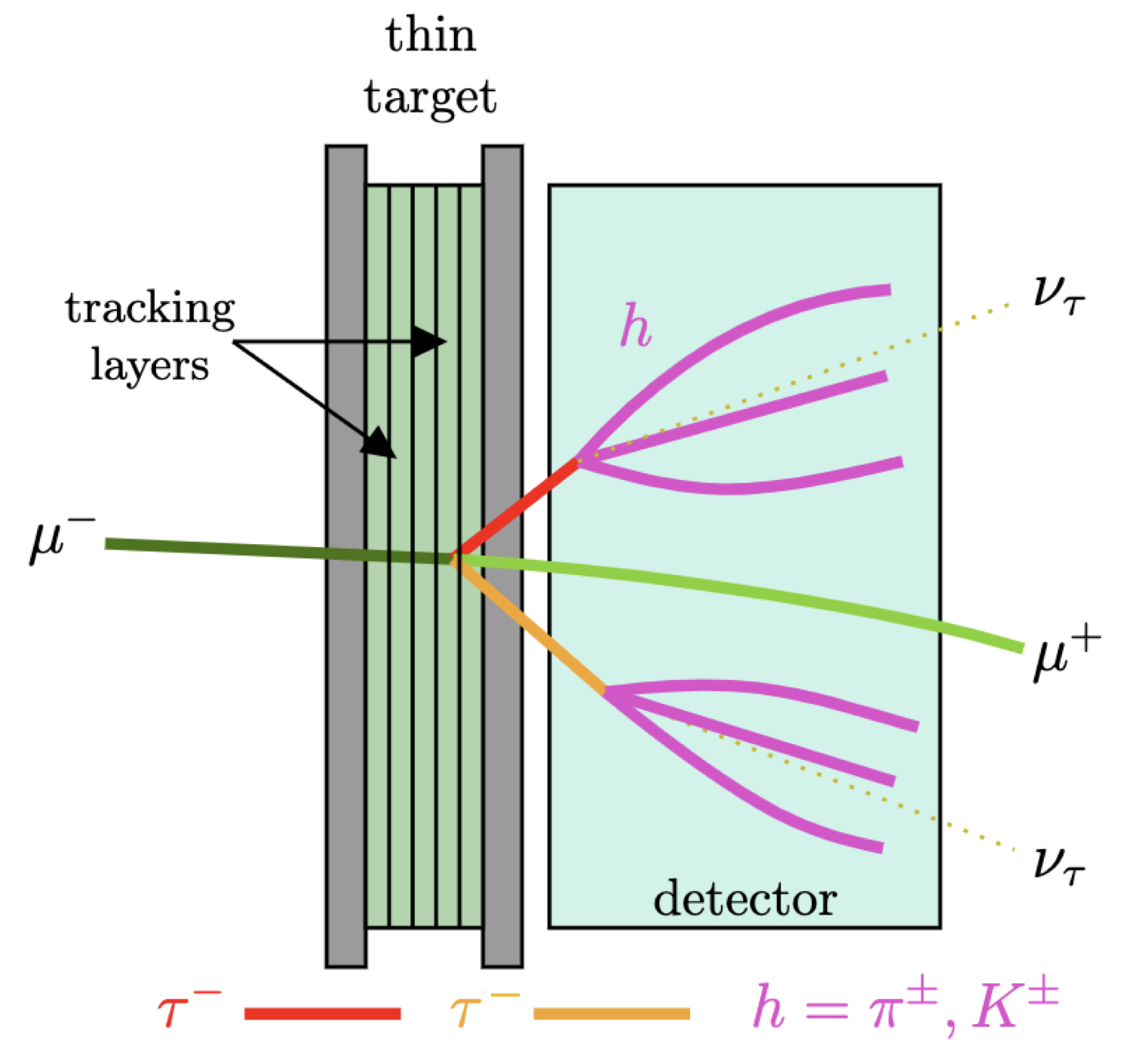}
    \caption{The LFV signature of the ALP $\mu\tau$ coupling in a thin metal plate (light green rectangle on the left) surrounded by electronic detectors (gray bars) and placed in front of a spectrometer (light blue) at a high-energy muon beam-dump experiment. The incident negative muon interacts coherently with a heavy nucleus in the target material, producing the ALP and negative tau lepton. The thin target material could be additionally interleaved with tracking layers, as indicated by black vertical lines in the plot. In the considered process, the ALP immediately decays into DM or the tau and muon pair. We present the latter case in the figure. The outgoing muon can have the opposite charge to the incident one, which we illustrate by varying the green shade of both lines. The intermediate tau lepton tracks are shown in red and yellow. We considered the $3$-prong $\tau$ decay channel into three charged mesons ($h = \pi^\pm, K^\pm$) shown in purple and a neutrino (light brown dotted lines).}
    \label{fig:signature}
\end{figure}

Here we propose a complementary active thin target concept for a high energy muon fixed-target experiment, schematically illustrated in \cref{fig:signature}. For concreteness, we consider a thin target composed of a $2~\textrm{cm}$ lead plate placed in between two veto or tracking layers. The target is positioned directly in front of a spectrometer. Interleaving this target with additional tracking layers could further improve the physics sensitivity. In the following, we will assume that a $2~\textrm{mm}$ track identification resolution can be achieved in the lead target material using electronic detectors or emulsion. We stress that we consider a much smaller total detector length than FASER$\nu$2, which facilitates such an interleaved design. 

The inclusion of interleaved tracking layers will allow for improved spatial resolution and event timing capabilities. Importantly, however, even the simplest detector setup in which the $2~\textrm{cm}$ lead plate is not instrumented inside will be sufficient to probe interesting regions of the parameter space of the ALP models. The interleaved tracking layers should then be treated as optional for this search.

Secondary visible particles produced in high-energy muon interactions in the active material will be measured in the detector placed downstream. This could follow the design of the aforementioned detector of a beam-dump experiment. On top of tracking capabilities and energy measurement, muon identification is essential to study the signal of our interest. In particular, we assume that  $\mu^\pm$s can be disentangled from charged pions. 
The active thin target run could operate concurrently with the beam-dump mode. In this case, even a small fraction of all of the muons of order $10^{16}$ MOT is enough for a meaningful search, as we discuss below. We also note that the muon beam should be largely unaffected by the active thin target detector, suggesting that our experiment could be placed upstream of and run in tandem with a thick target muon beam dump experiment, such as the setup proposed in Ref.~\cite{Cesarotti:2022ttv}.

For reference, $10^{16}$ MOT corresponds to only about an hour of operation of the muon beam source to be designed for the muon collider, assuming $5~\textrm{Hz}$ muon source repetition rate and $2\times 10^{12}$ muons per bunch relevant for the MAP parameters~\cite{Delahaye:2013jla,Neuffer:2018yof}. We assume these muons will be dumped on the active target material before decaying. Below, we will present the results for a $1.5~\textrm{TeV}$ muon beam. We reiterate that it is conceivable to assume a thin-target detector that we propose to operate in the preliminary stage of the work toward constructing the muon collider, as it relies on only a single high-energy muon beam.

In particular, using such a muon beam with $1~\textrm{TeV}$ energy has recently been investigated in the context of the proposed $\mu$TRISTAN project at the J-PARC facility~\cite{Hamada:2022mua}. While the proposed concept focuses on colliding muon and electron beams, the search discussed in our study could rely on installing an additional dedicated fixed-target experiment to capture muons leaving the accelerator before decaying to study LFV ALPs and other feebly-interacting particles.

We also emphasize that the intensity and energy of the muon beam considered here are unprecedented in current and near-future experiments, which should be considered when designing the experiment. For instance, a highly collimated muon beam could potentially damage the active detector. In this case, beam defocusing would be advantageous before it hits the target. The experiment could also operate under a reduced beam intensity in a short, dedicated run. Additional effects will be related to interactions of an electron cloud surrounding the muon beam. Being less energetic than muons, these could be partially deflected away or shielded on their way to the detector. These and other detector requirements, e.g., cooling, must be analyzed in detail to assess the feasibility of such a detector. We leave this for future studies, while, in the following, we focus on the discovery prospects for ALPs with LFV couplings.

\subsection{Signal and Backgrounds\label{sec:signalBG}}

Having described the experimental facilities of interest, we now discuss details of the signal selection process and expected backgrounds. These depend on the ALP decay mode.

\medskip
\textbf{Invisibly decaying ALP:} In the case of an invisibly decaying ALP, much of the incident muon energy is carried by the massive pseudoscalar particle in the scattering process $\mu^\pm N \to \tau^\pm N (a^\ast\to\chi\chi)$, leading to substantial missing energy. Instead, the outgoing tau lepton decays promptly into semi-visible final states. The signal selection prospects depend on the $\tau$ decay mode and the capability to reconstruct the intermediate short tau track before it decays.

The tau lepton can decay leptonically to a muon with a branching fraction of $\mathcal{B}(\tau^-\to\mu^-\bar{\nu}_\mu\nu_\tau) = 0.1739$~\cite{ParticleDataGroup:2022pth}. In this case, and for small tau decay lengths, the only detectable tracks are the ones left by the incident and outgoing muons, with a significant invisible energy loss and the muon kink. 

\underline{NA64-$\mu$} The relevant muon missing momentum signature is currently studied in the NA64-$\mu$ experiment~\cite{Sieber:2021fue}. In this case, we assume that the intermediate tau lepton track cannot be reconstructed and we focus on only the muon tracks. In the search, the outgoing muon is required to carry $<80~\textrm{GeV}$ energy, i.e., less than half of the incident $160~\textrm{GeV}$ beam energy. We have verified that this is the case for our events of interest, given the significant missing energy associated with the produced ALP with $m_a\gg m_\mu$ and neutrinos from the tau lepton decay. Beyond this, events with energy deposition in the ECAL that is larger than that expected from minimum ionizing particles are rejected. Similarly, events with more than one charged track in the tracking detectors or additional activity in the veto and hadronic calorimeters are also rejected. These cuts are satisfied for coherent scatterings, while diffractive processes can also satisfy them, especially for low-energy proton recoils. In the following, we assume $100\%$ detection efficiency in the signal region to ensure a fair comparison with other experiments, cf. Ref.~\cite{Andreev:2024sgn} for the detailed discussion about the signal selection and background rejection in NA64-$\mu$.

\underline{FASER$\nu$2} A similar muon missing-momentum signature was analyzed for the FASER$\nu$2 detector in the FPF in Ref.~\cite{Ariga:2023fjg}. As discussed therein, the excellent capabilities of emulsion detectors could allow for the identification of the interaction vertex and reconstruction of the kink angle with high precision. The measurement of the incident muon energy would rely on multiple Coulomb scatterings in FASER$\nu$2, while the outgoing muon energy and charge could be additionally measured with the FASER2 spectrometer placed downstream of the emulsion detector. The presence of the final-state muon is essential to guarantee that the interface tracker placed in between the emulsion detector and spectrometer is activated. This helps properly triggering such events, which otherwise remains much more challenging due to the lack of time stamps in the emulsion alone. 

It is important to mention that this search in FASER$\nu$2 could suffer from substantial backgrounds from the muon-induced photon bremsstrahlung process. A priori, given the lack of time information, it remains challenging to properly associate the photon-induced $e^+e^-$ pair production vertex in a separate part of the detector with the parent muon kink. Ref.~\cite{Ariga:2023fjg} discusses possible improvements in track reconstruction and event detection algorithms that would be necessary to perform this search in emulsion.

\begin{figure*}[t]
    \centering
    \includegraphics[width = \linewidth]{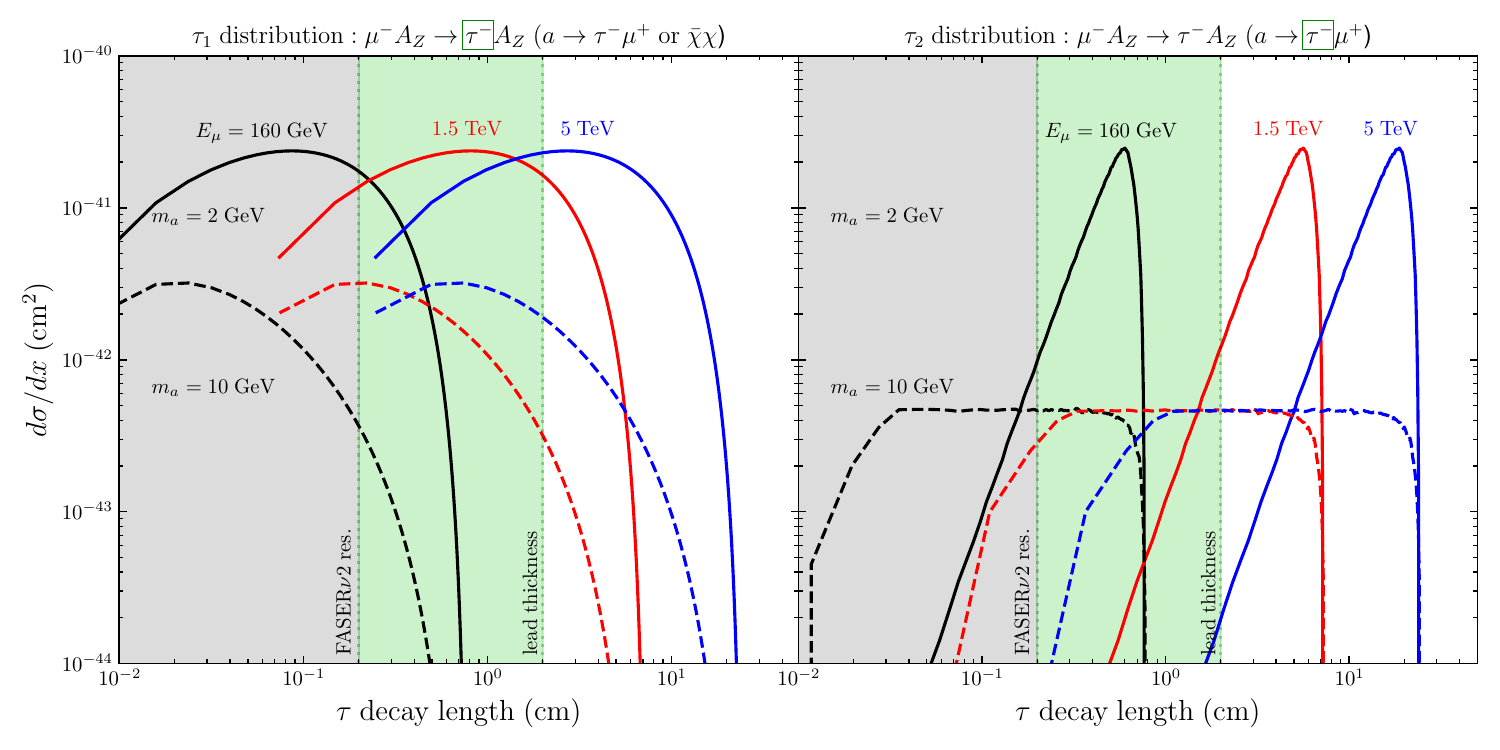}
    \caption{\textit{Left}: The differential cross section distribution for the ALP production process, $\mu A_Z \to \tau a A_Z$, where $A_Z$ denotes a nucleus $N$ of charge $Z$ (here we consider lead, $Z_{\rm Pb} = 82)$, with respect to the energy transfer to the tau lepton. The distribution is shown as a function of the tau lepton decay length. The black, red, and blue solid (dashed) lines correspond to incident muon energies of $160~\textrm{GeV}$, $1.5~\textrm{TeV}$, and $5~\textrm{TeV}$, respectively, for an ALP mass $m_a = 2~\textrm{GeV}$ ($10~\textrm{GeV}$). The gray-shaded region on the left corresponds to the expected emulsion detector resolution of $2~\textrm{mm}$, which we also assumed for the high-energy muon fixed-target detector instrumented with additional tracking layers. The green-shaded region corresponds to $2~\textrm{cm}$, which is the considered thickness of the lead target plate at the future fixed-target experiment.  \textit{Right}: Similar to the left but for the tau lepton produced in the subsequent process of the visible ALP decay, $a\to \mu\tau$. 
    }
    \label{fig:taudistribution}
\end{figure*}

On the other hand, if the intermediate tau lepton track is detected, the bremsstrahlung backgrounds could be greatly suppressed. In this case, two consecutive bremsstrahlung processes with a high muon energy loss and a small distance between the interaction vertices would have to happen to mimic the signature. The background rate is drastically reduced then, assuming that both muon kinks can be resolved separately. 

In the left panel of \cref{fig:taudistribution}, we show the decay length distribution of the $\tau$ lepton produced in the incident muon LFV scattering process, $\mu N\to \tau a N$. We present the results for several initial muon energies, $E_\mu = 160~\textrm{GeV}$, $1.5~\textrm{TeV}$, and $5~\textrm{TeV}$, and for two representative ALP masses, $m_a = 2~\textrm{GeV}$ and $10~\textrm{GeV}$. The $y$ axis of the plot corresponds to the differential scattering cross section for the ALP production in muon scatterings off the heavy nuclei, and $x = 1 - E_a/E_\mu \simeq E_\tau/E_\mu$ is the relative muon energy transfer to the tau lepton. In the plot, we also mark the aforementioned expected track reconstruction resolution in the emulsion detector, $\delta_{\textrm{FASER$\nu$2}} = 2~\textrm{mm}$, as a gray-shaded region. As one can deduce from the plot, for $m_{a} = 2~\textrm{GeV}$ the incident muons with $E_\mu$ of order a few hundred GeV or so will produce tau leptons with a sufficiently large decay length to be detected in FASER$\nu$2 as a separate short track. 

The precise detector resolution will additionally depend on the kink angles, as extremely collimated tracks are more difficult to disentangle. Since both the ALP production vertex and the subsequent tau decay would produce significant missing energy and momentum, we assume for simplicity that such a reconstruction can be performed thanks to the  impressive angular resolution of emulsion detectors if the distance between the vertices exceeds $\delta_{\textrm{FASER$\nu$2}}$. The expected single-layer angular resolution in FASER$\nu$ is $\sigma_{\textrm{ang}}\sim 0.23~\textrm{mrad}$~\cite{Ariga:2023fjg}. Detailed detector simulations should further verify this assumption, which is beyond the scope of this work.

In the following, we present the projected FASER$\nu$2 bounds after requiring in our simulations that the $\tau$ decay length is larger than the track identification resolution. Hence, we require that the entire chain of $\mu\to\tau\to\mu$ tracks can be reconstructed, and there are no additional charged tracks emerging from the two interaction vertices of interest. We then assume backgrounds can be suppressed to a negligible level. For completeness, we also show the ultimate sensitivity of FASER$\nu$2 after considering all the $\tau$ decay modes and no additional cuts.

\underline{High-energy muon fixed-target experiment} The proposed experiment employing a high-energy muon beam could use a similar detection strategy. In this case, we consider a fixed-target experiment with only a 2-cm lead plate used as a thin target. After correcting for the difference in the detector size and inverse interaction length in lead and tungsten, the ALP production probability is more than two orders of magnitude lower in this plate than in FASER$\nu$2. However, the expected sensitivity of the detector can be driven by the large number of MOT that could exceed the FPF predictions by even a few orders of magnitude. As discussed above, we focus on one possible experimental setup with fixed muon energy equal to $E_\mu = 1.5~\textrm{TeV}$. This further enhances the production cross section for heavy ALPs compared with the broad muon energy spectrum at the FPF, which has a substantial fraction of muons at lower energies. 

Given the small total length of the considered detector, we assume it will be possible to time stamp events. This can be achieved thanks to the interleaved tracker layers and by measuring charged particles entering the detector after leaving the active thin target material. The timestamp and good position resolution could allow for better tau lepton identification by measuring its $3$-prong decays into three charged hadronic final-state particles (pions or kaons). The relevant branching fraction is equal to $\mathcal{B}(\tau_{\textrm{3-prong}}) \simeq 0.15$~\cite{ParticleDataGroup:2022pth}. We require this tau lepton decay mode when presenting results below.

\medskip
\textbf{Visibly decaying ALP:}
The search for ALPs decaying promptly and visibly into the $\mu\tau$ pair requires a different dedicated strategy. Combined with the ALP production, the relevant process produces one muon and two tau leptons in the final state, $\mu N\to \tau N(a\to\mu\tau)$. Depending on the lepton charges produced in the ALP decay, the final-state muon can have the same or the opposite charge as the incident one. The case with the conserved muon charge suffers from large SM backgrounds from di-tau production in high-energy muon scatterings with an intermediate photon or off-shell gauge boson emission~\cite{Bulmahn:2008fa}. We estimated the relevant number of SM events to be $N_{\textrm{SM},\mu\to\mu\tau\tau} \simeq 2.8$k in FASER$\nu$2, while it grows to $2.1\times 10^5$ in the $2$ cm thick lead plate in the future high-energy muon fixed-target experiment operating at $E_\mu = 1.5~\textrm{TeV}$ assuming $10^{16}$ MOT. 

\begin{figure*}
   \centering
   \includegraphics[width=0.49\linewidth]{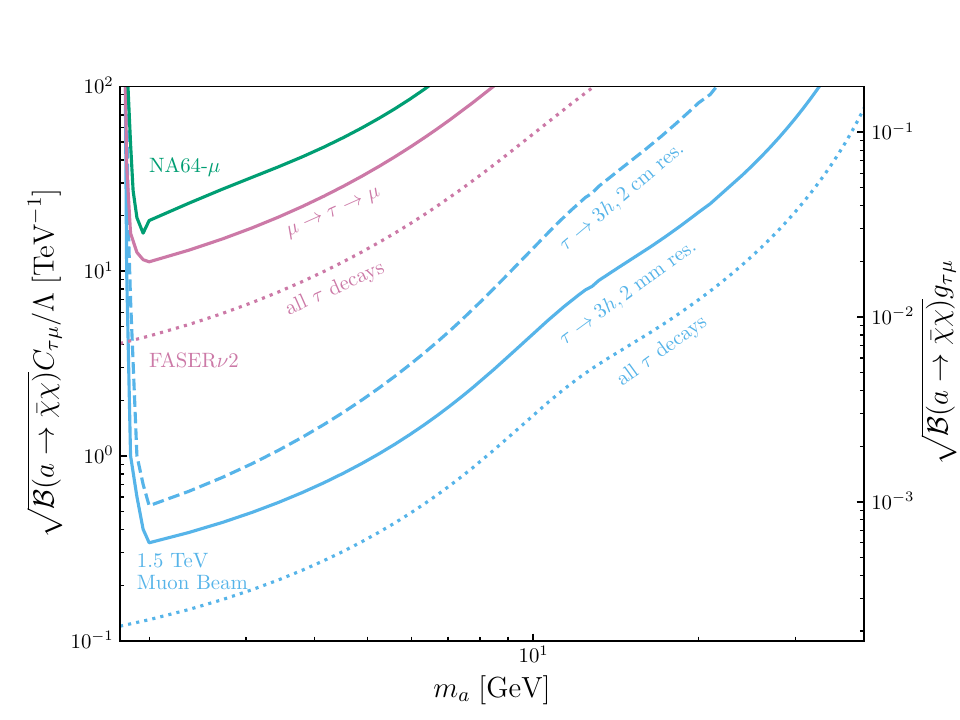}
    \includegraphics[width=0.49\linewidth]{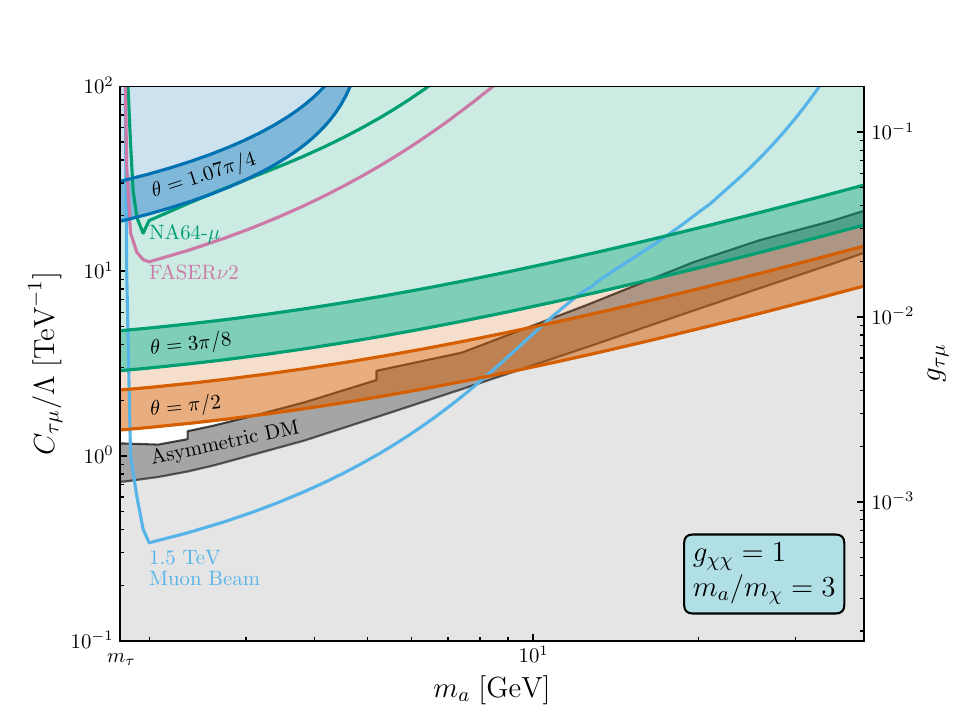}
   \caption{\textit{Left}: The projected exclusion bounds for the ALP model with a LFV $\mu\tau$ coupling as a function of the ALP mass $m_a$. We assume that the ALP decays invisibly into DM particles with a branching fraction $\mathcal{B}(a\to \chi\bar{\chi})$. The plot assumes zero background events after cuts and $N_{\textrm{ev}} = 3$ signal events. From top to bottom, the green solid line is the expected future sensitivity reach of the NA64-$\mu$ detector, the purple solid (dashed) line is the projected bound for FASER$\nu$2 with and without the cuts discussed in the text. The bottom blue lines correspond to the proposed high-energy muon fixed-target experiment with $E_\mu=1.5~\textrm{TeV}$ and $10^{16}$ MOT, for which we assume the outgoing tau lepton decays into the 3-prong final state. The dashed  line further assumes $\delta_{\textrm{MC}} = 2~\textrm{cm}$ track reconstruction resolution, the solid  line is obtained for $\delta_{\textrm{MC}} = 2~\textrm{mm}$, and the dotted line at the very bottom is the ultimate sensitivity derived with no additional cuts. \textit{Right}: Same as left panel but here we compare the projected exclusion bounds with the predictions for the $(g-2)_\mu$ anomaly and DM. We assume $g_\chi=1$ and $m_a = 3m_\chi$ in this plot. The dark colored bands on top correspond to the regions in the parameter space favored by $(g-2)_\mu$ for several selected values of the angle $\theta$, cf. \cref{sec:gm2}. The light colored regions above the corresponding colored band are disfavored for the given angle $\theta$ as the ALP contribution to $(g-2)_\mu$ is too large. The light gray-shaded region at the bottom is excluded as the predicted thermal DM relic abundance is too large. The dark gray-shaded band is excluded by DM ID and CMB bounds, cf. \cref{sec:DM}. In this case, we assume an asymmetric DM scenario with only a small thermal symmetric DM component sensitive to indirect searches. We discuss other bounds on this model in \cref{sec:otherbounds}.  
   }
   \label{fig:ALP_cuts}
\end{figure*}

Instead, the muon charge flip, e.g., $\mu^- N\to \tau^- \tau^-\mu^+ N$, marks the presence of explicit LFV and could be a smoking gun signature of BSM effects. We illustrate this in \cref{fig:signature} for an incident negative muon and $3$-prong tau lepton decays. Notably, this search requires charge measurement for both the incident and outgoing muon and capabilities to identify the two outgoing tau leptons. Charged hadrons produced in tau lepton decays should be further measured in a downstream hadronic calorimeter to disentangle them from the outgoing muon. Combined, such detection prospects would be possible to achieve in the future high-energy muon fixed-target experiment, and we focus on this proposed detector for further discussion of the ALP model with visible decays.  

\begin{figure}[t]
    \centering
   \includegraphics[width=\linewidth]{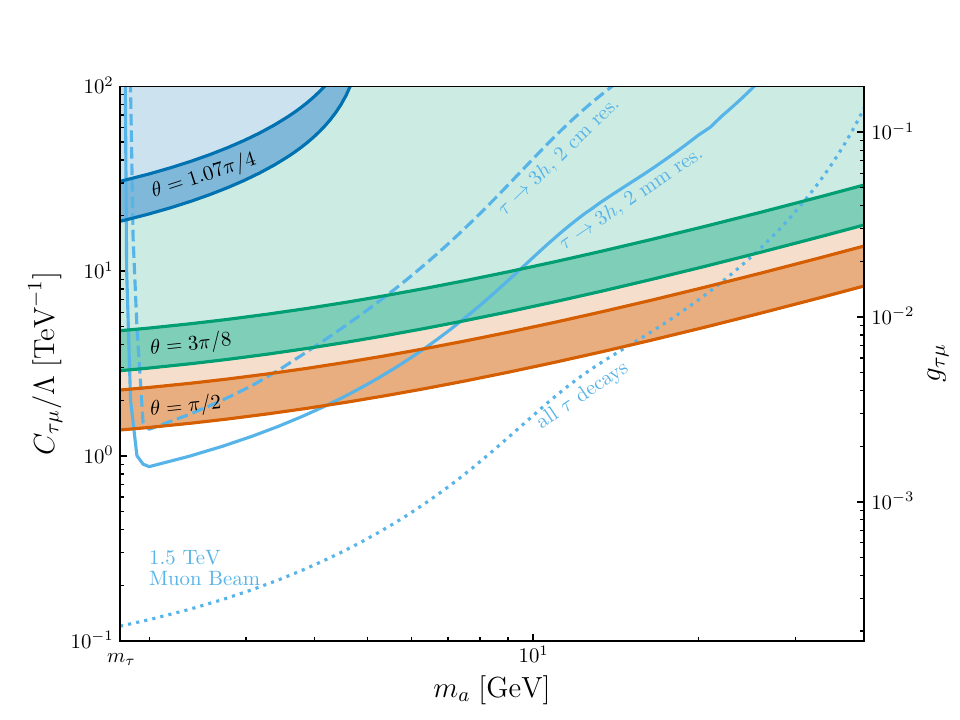}
    \caption{The same as the left panel of \cref{fig:ALP_cuts} but for the visibly decaying ALP, $a\to \mu^+\tau^-$, assuming an incident negative muon $\mu^-$, such that the muon charge flip can be observed. Here, only the future projected bounds for the high-energy muon fixed-target experiment are shown, and both tau leptons are required to decay into 3-prong final states. We also show colored bands in the parameter space, which are favored by the $(g-2)_\mu$ anomaly, assuming a few  selected values of the angle $\theta$.
    }
    \label{fig:ALP_visible}
\end{figure}

As discussed above, the tau lepton identification prospects can be greatly improved depending on the assumed detector spatial resolution. In the right panel of \cref{fig:taudistribution}, we show the $\tau$ decay length distribution corresponding to the tau lepton produced in visible ALP decays, $a\to\mu\tau$. As can be seen, for $m_a = 2~\textrm{GeV}$, the tau lepton will typically decay outside the lead plate, and $\delta_{\textrm{MC}} = 2~\textrm{cm}$ is sufficient to reconstruct such events properly. Improved detector resolution will, however, positively impact the tau identification for heavier ALPs, as shown for $m_a = 10~\textrm{GeV}$. In this case, the ALP carries away most of the incident muon energy and its energy is approximately fixed, $E_a\simeq E_\mu$. Since $m_a\gg m_\mu,m_\tau$ for the growing ALP mass, one expects a flat energy distribution of both the ALP decay products in the laboratory frame and we find a plateau in the distributions for $m_a = 10~\textrm{GeV}$. 

\section{Results\label{sec:results}}

\textbf{Invisibly decaying ALP} In the left panel of \cref{fig:ALP_cuts}, we illustrate the projected exclusion bounds of all considered experiments on the invisibly decaying ALP parameter space. In particular, we present the projected constraint for the NA64-$\mu$ experiment with $10^{13}$ MOT; we find that the current NA64-$\mu$ bounds exclude regions above $C_{\mu\tau}/\Lambda = 100~\textrm{TeV}^{-1}$, which is not shown in the plot. We present the projected FASER$\nu$2 bounds assuming that the entire chain of $\mu \to \tau \to \mu$ tracks can be reconstructed, as discussed in \cref{sec:signalBG}. We also show the ultimate FASER$\nu$2 sensitivity obtained, assuming all $\tau$ leptons can be properly reconstructed in the emulsion.

The plot also illustrates the impact of the cuts on the projected exclusion bounds of the high-energy muon fixed-target experiment with a $E_\mu = 1.5~\textrm{TeV}$ beam and $10^{16}$ MOT. In this case, we show the expected results for either $\delta_{\textrm{MC}} = 2~\textrm{mm}$ or $2~\textrm{cm}$ track reconstruction resolution. The better ($2~\textrm{mm}$) resolution requires using additional tracking layers interleaving the lead target, as discussed in \cref{sec:experiments}. The better the resolution $\delta_{\textrm{MC}}$, the stronger the expected bounds, which is more relevant for larger ALP masses as discussed above. For completeness, we also show the ultimate sensitivity of the high-energy muon fixed-target experiment obtained based on all possible tau decay modes and no additional cuts. 

In the right panel of \cref{fig:ALP_cuts} we compare our projected exclusion bounds for different experiments with the muon anomalous magnetic moment predictions. In particular, for several selected values of the angle $\theta$ (see discussion in \cref{sec:gm2}), we show the corresponding regions (dark colored bands) in which the $(g-2)_\mu$ theoretical prediction can be increased by the ALP contribution to match the experimental result within $2\sigma$. The regions above the dark colored bands are disfavored since the ALP contribution to $(g-2)_\mu$ is too large for the corresponding specified values of $\theta$. In particular, the ALP contribution to the anomalous magnetic moment of the muon is maximized for $\theta = \pi/2$, in which case the preferred region for $m_a\sim m_\tau$ is obtained for $C_{\mu\tau}/\Lambda \simeq \textrm{2}~\textrm{TeV}^{-1}$, i.e., for $g_{\mu\tau}\simeq 4\times 10^{-3}$. Other values of the angle then require larger ALP LFV couplings to account for the $(g-2)_\mu$ discrepancy. 

As can be seen, the expected bounds from NA64-$\mu$ and FASER$\nu$2 lie in parameter regions that are already excluded by $(g-2)_\mu$ for most values of the angle $\theta$, with the exception of $\theta\sim \pi/4$. Instead, much stronger projected exclusion limits can be obtained for the proposed high-energy muon fixed-target experiment. This will be able to thoroughly probe the remaining allowed region of the parameter space, in which the $(g-2)_\mu$ anomaly can be explained up to $\theta\sim \pi/2$ and $m_a\sim 10~\textrm{GeV}$ or so.

The scenario with the invisibly decaying ALP is additionally constrained by relic density and indirect detection bounds on asymmetric DM species, as discussed in \cref{sec:DM}. We show these bounds in the plot assuming $g_\chi = 1$ and $m_a = 3m_\chi$. As can be seen, these bounds and projected limits almost close the remaining parameter space of this model below the $(g-2)_\mu$ $2\sigma$ region obtained for $\theta=\pi/2$. These constraints limit the available parameter space from below, while the future high-energy muon fixed-target experiment could constrain it from above. Combined, these bounds will fully probe the asymmetric DM scenario for the considered benchmark model.

\medskip
\textbf{Visibly decaying ALP}
In \cref{fig:ALP_visible}, we present the projected exclusion bounds obtained for the visibly decaying ALP. As discussed above, in this case, we show the results only for the proposed future high-energy muon fixed-target experiment with $10^{16}$ MOT. From top to bottom, the exclusion limits correspond to the resolution $\delta_{\textrm{MC}} = 2~\textrm{cm}$ and $3$-prong tau decays, similar predictions for $\delta_{\textrm{MC}} = 2~\textrm{mm}$, and the ultimate sensitivity with no cuts. In the $3$-prong decay case, we require both tau leptons to decay into this final state, so the projected exclusion bounds are slightly worse than in the invisibly decaying ALP scenario. The visible decay bounds additionally correspond to the muon charge flip signature, i.e., we require the ALP to decay into the opposite charge muon compared to the incident one. 

Despite these limitations, values of the coupling constant as low as $g_{\mu\tau}\sim 10^{-3}$ can be probed in this scenario. This also allows for a thorough test of the parameter space resolving the $(g-2)_\mu$ discrepancy. As discussed in \cref{sec:DM}, the correct DM abundance, in this case, can be obtained thanks to secluded annihilations, $\chi\bar{\chi}\to a a$, independent of the ALP LFV coupling. In particular, assuming $m_\chi = 3m_a$, this can be obtained for $g_\chi \sim 0.1\times \sqrt{m_a/\textrm{GeV}}$. Hence, for visibly decaying ALPs, the DM relic target can also be probed by the future high-energy muon fixed-target experiment assuming an $\mathcal{O}(1)$ value of the dark coupling constant. We note that the active thin target detector will provide an independent probe of this scenario, complementary to future constraints from central detectors at high-energy colliders discussed in \cref{sec:otherbounds}, and it can extend beyond these projected bounds if the incident muon number can be increased beyond $10^{16}$ MOT as considered in our study.

\section{Conclusions\label{sec:conclusions}}

ALPs constitute a well-motivated class of light BSM physics and may play a key role in addressing some of the open questions in particle physics and cosmology. 
In particular, new ALP interactions in the charged lepton sector are motivated by unresolved experimental anomalies such as the muon anomalous magnetic moment, and, furthermore, such ALPs may serve as a mediator between the SM and DM. It is therefore of great interest to search for new phenomena associated with ALP-charged lepton interactions, including those with LFV.  

In this work, we have studied the discovery prospects for one such scenario in which ALPs with $\mathcal{O}(10~\textrm{GeV})$ masses and LFV couplings to the $2$nd and $3$rd generation charged leptons mediate interactions between the dark and visible sectors. As we have demonstrated, there is substantial open parameter space in which both the $(g-2)_\mu$ anomaly can be explained and the observed DM abundance can be produced in the early universe. For a Dirac fermion DM candidate, the latter requires assuming an initial asymmetry in the DM sector for $m_\chi < m_a$, or alternatively it can be achieved through standard thermal freezeout of secluded annihilations of heavier DM to lighter ALPs.

While searches for such LFV ALP mediators in this mass range remain challenging at lower energy experiments, and with experiments using electron or proton beams, they become more promising in setups utilizing high-energy muon beams. In particular, we have obtained projected exclusion bounds for the ongoing NA64-$\mu$ experiment and the proposed FASER$\nu$2 detector at CERN. We have also proposed an active muon beam thin-target experimental concept for a future high-energy muon facility, which could be carried out as a part of an R\&D phase of the preparations for a future muon collider. We have demonstrated that such an experiment has the potential to thoroughly probe a significant portion of the $(g-2)_\mu$ and DM motivated parameter space of the model for ALP masses up to tens of GeV. The proposed active thin-target detector can also be used to search for other new physics species coupled to muons and we leave this for future studies.

\vspace{0.5cm}

{\tt Digital data for this work can be found as  ancillary files in the arXiv submission~\cite{data}.}

\section*{Acknowledgements}
We thank Matt Forslund for discussions and Felix Kling for comments on the manuscript. S.T.  would like to thank Mary Hall Reno for sharing her numerical code used to estimate the muon-induced di-tau production in the Standard Model~\cite{Bulmahn:2008fa}.
The work of B.B. is supported by the U.S. Department of Energy under grant No. DE–SC0007914.
The work of H.D. is supported by the U.S. Department of Energy under
Grant Contract DE-SC0012704.  The work of E.~N.~ and R.~M.~is supported by the U.~S.~Department of Energy under Grant Contract DE-SC0010005.  This material is based upon work supported by the U.~S.~Department of Energy, Office of Science, Office of Workforce Development for Teachers and Scientists, Office of Science Graduate Student Research (SCGSR) program. The SCGSR program is administered by the Oak Ridge Institute for Science and Education for the DOE under contract number DE‐SC0014664.  S.T. is supported by the National Science Centre, Poland, research grant No. 2021/42/E/ST2/00031.

\begin{raggedright}
\bibliography{LFV-FASERnu}
\end{raggedright}

\onecolumngrid

\appendix

\section{Example UV completion} \label{app:UV}

In this section, we outline a simple UV model that can give rise to the types and magnitudes of couplings that are explored in this work.  This model is not meant to be a complete proposal for understanding flavor physics, but rather as a simple proof of principle.

We introduce new left- and right-handed Weyl fermion fields $F_{L,R}$ that have the same quantum numbers as the right-handed leptons of the SM.  Here, $F_L$ has charge $Q_X=+1$ under a new global $U(1)_X$ symmetry.  In addition, we include two complex scalar fields $\Phi_1$ and $\Phi_2$, both also with charge $Q_X=+1$ under the new $U(1)_X$.  None of the SM fields carry $U(1)_X$ charge.  A summary of the new fields and their charges is given in \cref{tab:uv_fields}.
\begin{table}[h]
\begin{tabular}{|c|cccc|}
\hline
&$U(1)_X$&$U(1)_Y$&$U(1)_{L_\mu}$&$U(1)_{L_\tau}$\\
\hline
$F_L$ & +1 & -1 & +1 & 0 \\
$F_R$ & 0 & -1 & 0 & +1 \\
$\Phi_1$ & +1 & 0 & 0 & 0 \\
$\Phi_2$ & +1 & 0 & 0 & 0 \\
\hline
\end{tabular}
\caption{New fields in the UV model and their charges under various U$(1)$ symmetries.  U$(1)_X$ is a new global symmetry, U$(1)_Y$ is hypercharge, and U$(1)_{L_\mu}$ and U$(1)_{L_\tau}$ are the lepton numbers associated with muons and $\tau$'s in the SM. As discussed in the text, the Weyl fermions $F_L$ and $F_R$ have the same quantum numbers as the right-handed SM muon and $\tau$ respectively.  \label{tab:uv_fields}}
\end{table}
In general, two complex scalar fields $\Phi_1$ and $\Phi_2$ will conserve two distinct $U(1)$ symmetries, but we presume that one linear combination is broken explicitly (for example by an interaction of the form $(\Phi_1^\dagger)^2 \Phi_2^2$ in the scalar potential), leaving only $U(1)_X$ to be spontaneously broken by the vevs of $\Phi_{1,2}$ and give rise to a single massless Goldstone boson which we call $a$.  Further soft breaking of $U(1)_X$ will give mass to $a$.

Adding the heavy fermions $F$ as well as the SM lepton fields, we have the following interaction terms:
\begin{equation}
{\cal L} = y_1 \Phi_1 \bar F_L F_R + y_2 \Phi_2 \bar F_L \mu_R + \lambda H \bar F_R \ell_\tau + \text{\small H.C.}\,,
\label{eq:L_UV}
\end{equation}
where $\ell_\tau$ is the left-handed $\tau$ doublet.  The interaction $y_1$ explicitly breaks the two individual lepton number symmetries down to the diagonal combination $U(1)_{L_{\mu} + L_{\tau}}$.  Most other Yukawa terms are forbidden by $U(1)_X$ or by this residual lepton-number symmetry; for the other remaining interactions, we assume their corresponding Yukawa couplings are small enough that they can be neglected. 

This model is free of gauge anomalies.  There is a global anomaly between $U(1)_X$ and $U(1)_Y$, which can lead to an $\mathcal{O}(\alpha)$ coupling of the form $a F_{\mu \nu} \tilde F^{\mu \nu}$.  We neglect this $a\gamma\gamma$ coupling in our main analysis, but the coupling generated by the global anomaly is likely small enough to not give significant new bounds; at any rate, it could be reduced by further modifications to this simple UV model if necessary.

We assume the form of the scalar potentials for the $\Phi_1$ and $\Phi_2$ fields is such that both obtain non-zero vevs, $\vev{\Phi_1} = v_1/\sqrt{2}$ and $\vev{\Phi_2} = v_2/\sqrt{2}$, both of which break the $U(1)_X$ symmetry.  The Higgs field also acquires a vev $\vev{H} = v/ \sqrt{2}$, with $v \approx 246$ GeV.  Inserting these vevs into \cref{eq:L_UV} gives rise to three parameters with dimensions of mass: 
\begin{equation}
M_F \equiv y_1 v_1 / \sqrt{2}; \ \ 
m_{\mu F} \equiv y_2 v_2/ \sqrt{2}; \ \ 
m_{\tau F} \equiv \lambda v/\sqrt{2}.
\end{equation}
We will consider the scenario in which $M_F \gg m_{\mu F}, m_{\tau F}$, and further will take the Yukawa coupling $\lambda \sim 1$, which means that $m_{\tau F} \sim v$.  After diagonalization of the mass matrix, the $\tau$ and $\mu$ leptons mix slightly with the Dirac fermion $F$, and their masses are also changed from their SM values, albeit very slightly; the corrections to the $\mu$ and $\tau$ masses are of order $1/M_F^2$.

To continue, we parameterize each of the complex scalars in terms of its magnitude and phase, $\Phi_i \equiv \tfrac{1}{\sqrt{2}} (s_i + v_i) e^{ia_i / v_i}$.  We take the scalar modes $s_i$ to be very heavy, and also assume that the single light Goldstone mode $a$ can be identified with the pseudoscalar $a_2$ associated with $\Phi_2$.  We assume that a soft mass $m_a$ is generated for $a$.  Integrating out the heavy $F$ fermion, the Lagrangian then contains the flavor-changing interaction of interest:
\begin{equation}
\mathcal{L} \supset -\frac{i m_{\mu F} m_{\tau F}}{M_F v_2} a \bar{\tau} \left( \frac{1+\gamma^5}{2} \right) \mu.
\end{equation}
Comparing this to \cref{eq:L_mutau_simple}, we find for the effective theory $\theta = \pi/4$ and (taking $m_\mu \ll m_\tau$),
\begin{equation}
\frac{C_{\mu \tau}}{\Lambda} = \frac{1}{M_F} \frac{m_{\mu F} m_{\tau F}}{v_2 m_\tau} = \frac{1}{M_F} y_2 \lambda \frac{v}{m_\tau}.
\end{equation}
If we take $M_F \approx 10$ TeV and $m_{\mu F} \sim v_2$ (so that the ratio $m_{\mu F}/v_2 \sim 1$), then we have
\begin{equation}
\frac{C_{\mu \tau}}{\Lambda} \approx 14\ {\rm TeV}^{-1}.
\end{equation}
Modest adjustments to the parameters can easily yield a value of $C_{\mu \tau} /\Lambda \sim 10^2\ {\rm TeV}^{-1}$, the largest values that we consider in this work, without resulting in any inconsistencies in the UV model. 

Other couplings are generated by the UV model, in particular new couplings of the Higgs boson.  As a consequence of the mass diagonalization between $\mu$, $\tau$ and $F$, the Lagrangian will contain the term
\begin{equation}
\mathcal{L} \supset \frac{\lambda m_{\mu F}}{\sqrt{2} M_F} h (\bar{\tau}_L \mu_R + \bar{\mu}_R \tau_L),
\end{equation}
which can mediate the flavor-violating decay $h \rightarrow \mu \tau$.  The resulting decay width for the Higgs would be 
\begin{equation}
\Gamma(h \rightarrow \tau \mu) \approx \frac{m_h}{16\pi} \frac{\lambda^2 m_{\mu F}^2}{M_F^2},
\end{equation}
with the approximation from taking $m_\mu, m_\tau \ll m_h$.  From Ref.~\cite{CMS:2021rsq}, the limit on the branching fraction is $\mathcal{B}(h \rightarrow \tau \mu) < 0.15$, which we can rewrite in the form
\begin{equation}
\frac{\lambda^2 m_{\mu F}^2}{M_F^2} = \frac{C_{\mu \tau}^2}{\Lambda^2} \frac{v_2^2 m_\tau^2}{v^2} \lesssim 2 \times 10^{-6},
\end{equation}
taking the Higgs width $\Gamma_h = 4$ MeV.
With the numerical values we have assumed above of $\lambda = 1$ and $M_F \approx 10$ TeV, this becomes $m_{\mu F} \sim v_2 \lesssim 14$ GeV. Thus, for relatively large $C_{\mu \tau}/\Lambda$, we may anticipate a correlated $h \rightarrow \tau \mu$ signal in this UV completion. 

Finally, additional constraints can arise from decays of the form $h \rightarrow a a$, with the ALPs decaying subsequently to $\mu \tau$.  This was studied in Ref.~\cite{Davoudiasl:2021haa}. Since our UV model preserves an approximate shift symmetry of the $a$ field the leading operators involving $\partial_\mu a$ will be generated (the equations of motion for $a$ were used to write the coupling $a \bar{\tau} \mu$ above).  A box diagram involving the fermions $\mu$, $\tau$ and $F$ will, after integrating out the heavy $F$ fermions, generate a coupling of the form
\begin{equation}
\mathcal{L}_{\rm eff} \supset \frac{1}{M_F^2} (\partial_\mu a)^2 H^\dagger H \rightarrow \frac{v}{M_F^2} h (\partial_\mu a)^2,
\end{equation}
which, in the notation of Ref.~\cite{Davoudiasl:2021haa} gives
\begin{equation}
\frac{C_{ah}}{\Lambda^2} \sim \frac{1}{M_F^2}.
\end{equation}
The constraints from searches studied in Ref.~\cite{Davoudiasl:2021haa} are only effective for $C_{ah} / \Lambda^2 \gtrsim 0.1 / {\rm TeV}^2$, whereas for $M_F \sim 10$ TeV we have $C_{ah} / \Lambda^2 \sim 10^{-2}/ {\rm TeV}^2$, well below the sensitivity of current searches.

\end{document}